\documentclass[12pt]{iopart}
\usepackage{float}
\usepackage{graphicx}
\usepackage{physics}
\usepackage[hidelinks]{hyperref}
\usepackage{cleveref}
\usepackage[dvipsnames]{xcolor}
\usepackage[backend=biber,style=numeric,citestyle=numeric-comp,sorting=none,doi=false,url=false,giveninits=true,maxnames=10]{biblatex}
\usepackage{multirow}

\DeclareFieldFormat
  [article,inbook,incollection,inproceedings,patent,thesis,unpublished]
  {title}{#1}

\renewbibmacro{in:}{%
  \ifentrytype{article}{%
  }{%
    \printtext{\bibstring{in}\intitlepunct}%
  }%
}

\DeclareFieldFormat{pages}{#1}
\DeclareFieldFormat[article]{number}{\mkbibparens{#1}}

\renewbibmacro*{volume+number+eid}{%
  \setunit{\addcomma\space}%
  \printfield{volume}%
  \printfield{number}%
  \setunit{\addcomma\space}%
  \printfield{eid}%
  \setunit{\addcolon}%
  \printfield{pages}}

\renewbibmacro*{issue+date}{%
  \setunit{\addcomma\space}%
  \iffieldundef{issue}
    {\usebibmacro{date}}
    {\printfield{issue}%
     \setunit*{\addspace}%
     \usebibmacro{date}}%
\newunit}

\renewbibmacro*{note+pages}{%
  \printfield{note}%
  \newunit}

\begin{document}

\title[]{Advancing the heralded photon-number-state characterization by understanding the interplay of experimental settings}

\author{Daniel Borrero Landazabal, Kaisa Laiho}
\address{German Aerospace Center (DLR e.V.), Institute of Quantum Technologies, Wilhelm-Runge-Str. 10, 89081 Ulm, Germany}
\ead{daniel.borrerolandazabal@dlr.de}

\begin{abstract}
We theoretically explore the properties of heralded number states including up to three photons that are generated from single-mode twin beams. We investigate the effects of different parameters involved in the state preparation by using the fidelity, normalized second-order factorial moment of photon number for the heralded state $(g^{(2)}_{\text{h}})$, and photon-number parity as figures of merit. Especially, the photon-number parity offers a practical and robust tool for inferring the target state quality by capturing the contamination of all undesired photon-number contributions. We focus on expressing our results in terms of experimentally easily accessible parameters such as the coincidences-to-accidentals ratio and the detection efficiencies. Our results identify the optimal parameter regions for generating high quality photon-number states by heralding and provide useful insights for advancing their use in quantum technologies.
\end{abstract}
\vspace{2pc}

\section{Introduction}

The generation and precise characterization of quantum optical states are foundational in order to advance the implementation of technologies for quantum information processing, quantum communication, and precision metrology \cite{migdall2013single}. For this purpose, single- and few-photon states are particularly interesting as they serve as indispensable resources \cite{kok2007linear}. These states can be produced for example by heralding on the detection of one of the twin beams generated in non-linear optical processes, such as parametric down-conversion (PDC) \cite{rarity1992quantum} or four-wave mixing \cite{levenson1985generation}. However, until today, the generation of high-quality photon-number states remains challenging due to both the experimental imperfections and the intrinsic properties of PDC of being a probabilistic process and of producing higher photon-number contributions, which introduces undesirable contributions to the photon statistics of the heralded state \cite{waks2006generation,laiho2009producing,sonoyama2024generation}.

The accurate characterization of the photon-number content of these heralded states is essential for the optimization and best exploitation of the photonic sources. Traditionally, the state quality has been accessed using the figures of merit such as the fidelity and the second-order correlation function, $g^{(2)}$ \cite{christ2012limits,francis2014exploring,laiho2019photon,meyer2020single,stasi2023high}. The fidelity quantifies the overlap of the generated state relative to an ideal target state, while $g^{(2)}$ provides information about the multi-photon contributions of the state. For single photons, a low $g^{(2)}$ value is desirable, as it signals minimal multi-photon contamination \cite{bruscino2024reduction,pani2024effects,rani2024bb84protocol}. However, these metrics present notable limitations: the fidelity is experimentally challenging to measure \cite{laiho2009producing}, and $g^{(2)}$ neglects the impact of the vacuum component, an essential factor in determining the overall state quality \cite{laiho2022measuring,zhang2024full,azuma2024heralded,magnoni2024toward}.

To address these limitations, we explore the photon-number parity as a more practical and comprehensive tool for characterizing photon-number states. The parity directly captures the quantum characteristics of the heralded state and provides insight into all photon-number contributions \cite{barnett2002methods,laiho2019photon} and eventually allows one to access the phase-space properties of light \cite{cahill1969density,royer1985measurement}. Its feasibility in an experimental implementation and robustness in evaluating state quality make it an attractive alternative to conventional metrics.

Photon correlations play, in general, an important role for the characterization of sources producing photons in pairs. In order to perform such correlation measurements with faint light, the use of photo detectors with single-photon sensitivity, high efficiency, low dead-time and low jitter are of great interest. Photon counters used in quantum optics experiments can mainly be divided in two categories. While the true photon-number resolving detectors allow implementing the projection into the photon-number basis, which enables a direct measurement of photon statistics and an easier treatment, the so-called quasi-photon-number resolving detectors, which can be implemented for example with beam-splitter networks connected to single-photon sensitive detectors that can only resolve between the optical vacuum and at least one impinging photon, can resemble a photon-number resolving detector \cite{santana2024extending}.

Here, we investigate the characteristics of heralded single-, two-, and three-photon states generated from single-mode twin beams in a PDC process. Our study evaluates the impact of experimentally easily accessible parameters on the quality of the generated states. We take into account the heralding efficiency and the heralded state detection efficiency that both can be accessed via the measurement of Klyshko's efficiencies \cite{klyshko1996observable} as well as the coincidences-to-accidentals ratio (CAR) that delivers us information of the PDC process strength being inversely proportional to the pumping power. By analyzing several figures of merit, including the fidelity, $g^{(2)}_{\text{h}}$-value, success probability and photon-number parity, we examine the trade-offs between state quality and practical experimental constraints. Our findings reveal that apart from the high detection efficiencies, the optimally selected CAR values indeed significantly enhance the quality of the heralded states. However, there is an inherent trade-off between the success probability and state quality, highlighting the need of the careful parameter optimization. Importantly, we demonstrate that the photon-number parity effectively captures these trade-offs and provides a reliable assessment of the state quality, surpassing the limitations of the fidelity and $g^{(2)}_{\text{h}}$-value. Altogether, we provide a deeper understanding of the practical challenges in generating high-quality photon number states by heralding and provide guidance for experimental implementations that are faced by the limitations and imperfections of the photon-counting applications.


\section{Methods used in heralding number states}
\label{sec:Methods}

The single-mode twin beams can be generated via the non-linear optical process of PDC \cite{lamas2001stimulated,fiorentino2007spontaneous,hou2016efficient,fang2024efficient,angural2024fully} or four-wave mixing \cite{levenson1985generation,embrey2015observation,yang2020multidimensional}. With a single-mode state, we refer to the presence of a single optical frequency mode that can be quantified by the Schmidt number \cite{christ2011probing}. We denote the twin beams as signal and idler, which  exhibit correlations across various degrees of freedom, including most importantly, the photon number and polarization \cite{rarity1992quantum,haderka2005direct,allevi2012measuring,dovrat2013direct,ALLEVI2022127828,paris2003remote}. The single-mode behavior can be achieved under specific conditions of the selected non-linear optical material platform as demonstrated experimentally in references~\cite{mosley2008conditional,evans2010bright,bruno2014pulsed,arora2024singly}. Typically, the individual twin beams are cross-polarized \cite{sempere2022reducing,sonoyama2024generation} or travel at different directions \cite{fiorentino2007spontaneous,lamas2001stimulated} such that one of them can easily be used for the heralding. 

Mathematically, the single-mode twin beams can be represented in the photon-number basis by \cite{barnett2002methods}
\begin{equation}
    \ket{\psi} = \sum_{n=0}^{\infty} \lambda^n \sqrt{1-|\lambda|^2} \ket{n,n}_{\text{s,i}}\, , 
\label{eq:smTWB}
\end{equation}
 where $\lambda$ denotes the squeezing strength parameter, $n$ represents the photon number, and $\text{s} (\text{i})$ labels the signal (idler) beam. This description assumes a perfect photon-number correlation between the signal and idler beams. The density matrix associated to the twin beams is then given by
 \begin{equation}
    \hat{\rho}_{\text{s,i}} =  \ket{\psi}\bra{\psi} = \left( 1-|\lambda|^2 \right)\sum_{n,m=0}^{\infty} \lambda^n (\lambda^*)^m  \ket{n,n}_{\text{s,i~i,s}} \bra{m,m} \, ,
\label{eq:DM_smTWS}
\end{equation}  
in which $^*$ denotes the complex conjugate.

Additionally, in the single-mode case, the relation between the twin beam mean photon number $\bar{n}$ and the squeezing strength, $|\lambda|^2 = \bar{n}/(1+\bar{n})$, allows one to express the joint photon-number statistics as
\begin{equation}
    P_{j,l} = |\braket{\psi}{j,l}_{\text{s,i}}|^2 \equiv P_{n=j=l}(\bar{n}) =  \frac{\bar{n}^{n}}{(1+\bar{n})^{n+1}}  \, , 
\label{eq:joint_statistics}
\end{equation}
incorporating the projection into the photon number basis spanned by the photon numbers $j$ and $l$ in signal and idler, respectively. Here $P_n(\bar{n})$ is the photon-number distribution, which follows in the single-mode case a thermal distribution given in terms of $\bar{n}$ \cite{stasi2023high,ALLEVI2022127828,paris2003remote}. As photons are generated in pairs within the non-linear optical medium \cite{haderka2005direct}, signal and idler beams are perfectly correlated in the photon-number basis. This gives null off-diagonal elements in the joint photon-number statistics of signal and idler, and provides both with equal mean photon number $\bar{n}$ \cite{barnett2002methods,paris2003quantum}. In the single-mode case, the thermal distribution $P_n(\bar{n})$ then determines the population of the different photon-number contributions that each depends on the twin beam mean photon number $\bar{n}$.  

\begin{figure}
    \centering
    \includegraphics[width=0.78\linewidth]{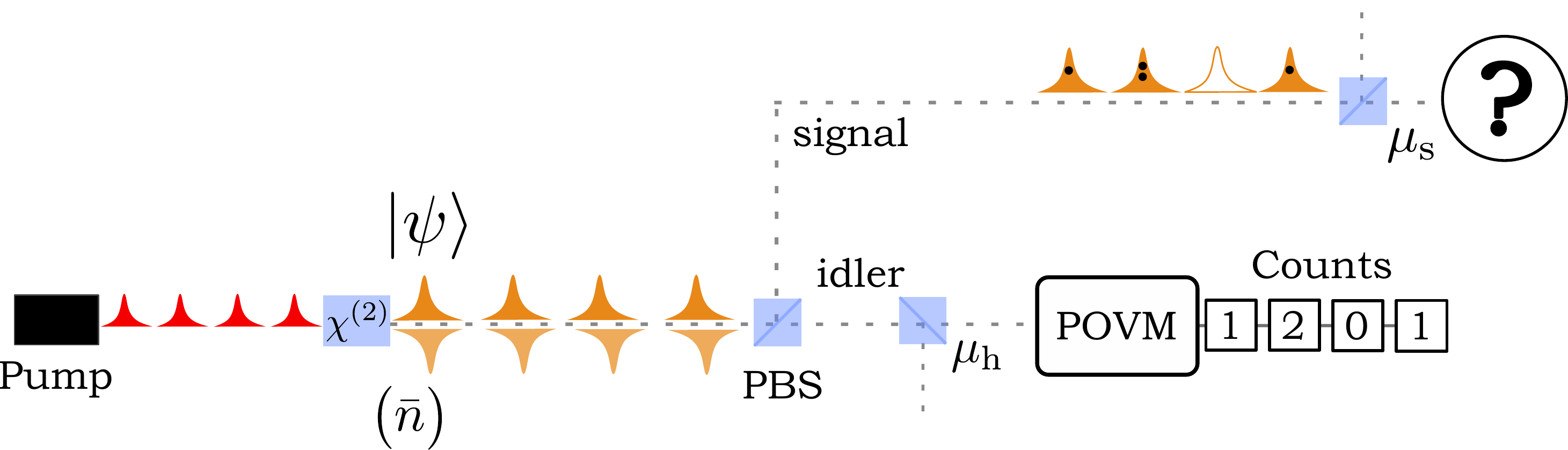} 
    \caption{Investigated experimental arrangement. Pump pulse splits to signal and idler beams in a type-II second-order nonlinear optical process. These emitted twin beams are described as single-mode twin beams given in equation~(\ref{eq:smTWB}), where the two correlated beams, signal and idler, are cross-polarized. The polarizing beamsplitter (PBS) is used to separate the two correlated beams. The idler pulses are detected by a photon counting system having an efficiency $\mu_{\text{h}}$ and an associated POVM, thus, the idler beam is used as a herald and a specific number of click count events realizes for each pulse. The heralded state in the signal arm travels through a lossy optical beam path having an efficiency $\mu_{\text{s}}$. In this example, the heralded 0-, 1- and 2-photon states are depicted with the corresponding number of black circles. Finally, the heralded state properties denoted with the symbol ``?" are characterized.}
\label{fig:Setup}
\end{figure}

Our primary objective is to investigate the characteristics of heralded single-, two-, and three-photon states generated by a source, whose properties follow equation~(\ref{eq:smTWB}). To achieve this, we examine the arrangement illustrated in figure~\ref{fig:Setup}. The generated twin beams are cross-polarized and can be thus separated by a polarizing beamsplitter, after which idler photons are detected with an efficiency $\mu_{\text{h}}$ using a photon-counting system associated with a typical Positive-Operator-Valued-Measure (POVM). Detection of one, two, or three photons in this photon counter heralds the single-, two-, and three-photon states in the signal arm, respectively. Once being heralded, the characterization of the state is carried out in the signal arm with a detection efficiency of $\mu_{\text{s}}$.

\subsection{Photon-counting system in the heralding}
\label{subsec:POVM}

Here, we employ the quasi-photon-number resolving detection system for the heralding process, such a system can be treated with the click-detection formalism \cite{sperling2012true}. In this case, the POVM \cite{scully1969quantum,barnett1998imperfect,lee2004towards,fiuravsek2025fundamental} associated with the photon counter used in heralding is an array of single-photon sensitive click-detectors without photon-number resolution. Thus, the POVM is defined as \cite{sperling2012true,sperling2014quantum}
\begin{equation}
    \hat{O}_k = :\binom{N}{k} \exp{\frac{-\mu_{\text{h}}\hat{n}+\nu}{N} (N-k)}\left( \hat{\mathcal{I}} - \exp{\frac{-\mu_{\text{h}}\hat{n}+\nu}{N}} \right)^k : \, ,
\label{eq:O}
\end{equation}
for an array of $N$ click-detectors, from which only $k=[0,N]$ are triggered when photons impinge on them. Additionally, the efficiency $\mu_{\text{h}}$ and dark count probability $\nu$ of the devices are taken into account. In equation~(\ref{eq:O}) $\hat{n}$ is the photon-number operator and the identity is denoted with $\hat{\mathcal{I}}$. Finally, the symbol $``::"$  refers to the normal operator ordering.

By expanding the last term in equation~(\ref{eq:O}) with binomial coefficients and applying the normal ordering, the investigated POVM takes  in the number-state basis the form
\begin{equation}
    \hat{O}_k = \sum_{m=0}^{k} \binom{N}{k}\binom{k}{m} (-1)^m e^{-\frac{\nu}{N}(N+m-k)}\sum^{\infty}_{n=0}\left(1-\frac{\mu_{\text{h}}}{N}(N+m-k)\right)^{n} \ket{n}_{\text{i~i}}\bra{n} \, ,
\label{eq:POVM}
\end{equation}
which spans in the Hilbert-space of the idler beam and only has diagonal elements. Such an operator assumes a uniformly distributed intensity over the click-detectors that are treated identically with each having dark count probability $\nu$ and efficiency $\mu_{\text{h}}$. For a more detailed calculation see appendix~\ref{appx:POVM}.

The detector parameters are of great importance, when heralding number states. For instance, the number of detectors can in principle be infinite, however, in a real-environment condition, practicality, space issues and reasonable costs play a relevant role. Therefore, we fix the number of detectors to $N=4$, which represents a realistic value achievable in a common experimental implementation. The number of clicks $k$ depends on the heralded target state, i.e., $k=1,2,3$ indicating a single-, two- and three-photon states, respectively. For the dark count probability a realistic value of $\nu=5\times 10^{-4}$ is taken from references~\cite{afek2009quantum,shibata2015ultimate,helt2017effect} providing a feasible range for the state of the art detectors. Such a value is also proper to show how the dark count probability affects the heralded state properties. Additionally, an estimation of its effect on the heralded single photon can be found in appendix~\ref{appx:darkcounts}.

In the event of heralding, the detection is performed in idler (see figure~\ref{fig:Setup}). Due to the photon-number correlation of twin beams this projection affects signal. Thus, the density matrix of signal upon heralding can be expressed as \cite{migdall2013single,van2017photodetector}
\begin{equation}
\begin{split}
    \hat{\rho}_{\text{s}} &= \frac{\Tr_{\text{i}}\left\{\hat{O}_k \hat{\rho}_{\text{s,i}}\right\}}{\Tr_{\text{s,i}}\left\{\hat{O}_k \hat{\rho}_{\text{s,i}}\right\}} \\
    &= \frac{\sum_{m=0}^{k} \binom{N}{k}\binom{k}{m} (-1)^m e^{-\frac{\nu}{N}(N+m-k)}\left(1-\frac{\mu_{\text{h}}}{N}(N+m-k)\right)^{n} P_n (\bar{n}) \ket{n}_{\text{s~s}}\bra{n}}{\sum_{n}^{\infty}\sum_{m=0}^{k} \binom{N}{k}\binom{k}{m} (-1)^m e^{-\frac{\nu}{N}(N+m-k)}\left(1-\frac{\mu_{\text{h}}}{N}(N+m-k)\right)^{n} P_n (\bar{n})} \, ,
\label{Eq:rho_signal}
\end{split}
\end{equation}
where $k$ is the number of heralded photons, $\hat{O}_k$ is the described POVM for the heralding and $P_n(\bar{n})$ is the thermal statistics of the single-mode twin beams.

\subsection{Photon statistics of heralded number states}

Assuming that the photon-number probability can be measured in a lossless fashion, its n-th element can be retrieved from equation~(\ref{Eq:rho_signal}) via
\begin{equation}
    p^{\text{s}}_n =\, _{\text{s}}\bra{n}\hat{\rho}_{\text{s}}\ket{n}_{\text{s}} = \Tr_{\text{s}}\left\{ \hat{\rho}_{\text{s}}\ket{n}_{\text{s}~\text{s}}\bra{n}\right\} \, .
\label{eq:heralded_statistics_1}
\end{equation}
By inserting equations~(\ref{eq:POVM}) and (\ref{eq:DM_smTWS}) into equation~(\ref{Eq:rho_signal}), and then plugging the result into equation~(\ref{eq:heralded_statistics_1}) one gets access to the photon-number distribution of the heralded signal beam including the effects of the imperfect heralding. This leads to
 \begin{equation}
    p^{\text{s}}_n = \mathcal{A} \sum_{m=0}^{k}\binom{N}{k}\binom{k}{m} (-1)^m e^{-\frac{\nu}{N}(N+m-k)}\left(1 -\frac{\mu_{\text{h}}}{N}(N+m-k)\right)^n P_n (\bar{n}) \, ,
\label{eq:heralded_statistics_2}
\end{equation}
in which $p^{\text{s}}_n$ is the photon-number distribution of the heralded signal beam when having $k-$clicks in the idler arm by using a system of $N$ click detectors with efficiency $\mu_{\text{h}}$ and  dark count probability $\nu$. Here $\mathcal{A}$ is the normalization constant so that $\sum_n p^{\text{s}}_n = 1$. 

The losses in the signal arm are treated by means of the loss-matrix evaluation \cite{laiho2009direct,afek2009quantum,migdall2013single,krishnaswamy2024retrieval}. By using the photon-number distribution of the heralded state in equation~(\ref{eq:heralded_statistics_2}), the loss-degraded photon-number distribution is evaluated as
\begin{equation}
    p^{\text{s}}_{m,L} = \sum_{n=0}^{N_{\text{s}}}\binom{n}{m} (\mu_{\text{s}})^m (1-\mu_{\text{s}})^{(n-m)} p^{\text{s}}_n \, ,
\label{eq:Lossy_statistics}
 \end{equation}
where $\mu_{\text{s}}$ and $N_{\text{s}}$ are the efficiency and the highest resolvable photon number of the detector system in the signal arm, respectively. Conversely, the inversion of equation~(\ref{eq:Lossy_statistics}) can be implemented to counteract the effect of the losses in the signal arm detection, which nevertheless is a highly nontrivial task in experiments \cite{krishnaswamy2024retrieval}.

\subsection{Defining twin beam properties via coincidences-to-accidentals ratio (CAR)}

In order to completely define the parameters---the input knobs---of our investigated experimental arrangement, we further need a parameter for characterizing the photon-number properties of the generated twin beams. In other words, we need access to the twin beam mean photon number $\bar{n}$ in equation~(\ref{eq:heralded_statistics_2}). Although being proportional to the pump power of the twin-beam generation, measuring such a value for faint light sources is not a straightforward task in a laboratory. This poses a challenge, as $\bar{n}$ is essential for completely describing the heralded state, as well as the probabilistic photon source itself \cite{stasi2023high,pani2024effects,rani2024bb84protocol,ALLEVI2022127828,bruscino2024reduction}. Luckily, for the single-mode twin beam, the cross-correlation between signal and idler provides a loss-independent alternative, which experimentally can be directly and easily measured \cite{christ2011probing,angural2024fully}. 

In this context, the cross-correlation $g^{(1,1)}$ represents the photon-number correlation between signal and idler that is averaged within the temporal length of a pulse \cite{laiho2022measuring,magnoni2024toward}. Such a correlation also corresponds to the broadly used coincidences-to-accidentals ratio (CAR) \cite{fang2024efficient,angural2024fully,fang2024high}. Its value can be computed for twin beams in terms of the photon-number content as
\begin{equation}
    \text{CAR}\equiv g^{(1,1)} =  \frac{\sum_{n=0}^{\infty}n^2P_{n}(\bar{n})}{\left(\sum_{n=0}^{\infty}nP_{n}(\bar{n})\right)^{2}} \, .
\label{eq:CAR_smTWS}
\end{equation}
For more details see appendix~\ref{appx:Cross-correlation_function}.

The value of CAR is related to the mean-photon number of the twin-beam state. Indeed, by inserting the thermal distribution, described in equation~(\ref{eq:joint_statistics}), into equation~(\ref{eq:CAR_smTWS}) delivers
\begin{equation}
    \text{CAR} = 2+1/\bar n \, .
\label{eq:approx_CAR}
\end{equation}
This applies especially only to the single-mode twin beams. Therefore, the CAR can be used as an alternative to the twin beam mean photon number and pump power.

Here, we focus on the effects of the efficiencies $\mu_{\text{h}}$, $\mu_{\text{s}}$ and the twin beam mean photon number---determined by the CAR---in the generation of heralded photon-number states. We vary these parameters to study the quality of the state preparation, by turning the input parameter knobs. Finally, the fixed and variable input parameter values used in the computation are summarized in table~\ref{Tab:parameters}.

\begin{table}
\centering
\begin{tabular}{c||ccc}
\multirow{ 2}{*}{Fixed} & Number of detectors & Number of clicks & Dark count probability \\
& $N=4$ & $k=[1,2,3]$  & $\nu=5\times 10^{-4}$  \\
\hline \vspace{-5mm} \\ \hline \vspace{-4mm}\\
\multirow{ 2}{*}{Variable} & Heralding efficiency & Signal efficiency &  Twin beam mean photon number \\ 
 & $\mu_{\text{h}}$ & $\mu_{\text{s}}$ & CAR\\
\end{tabular}
\caption{All fixed and variable input parameters for the computations. The number of clicks implies the heralding of a $k$-photon state.}
\label{Tab:parameters}
\end{table}


\section{Analysis tools for verifying and validating the heralded states' properties}
\label{sec:Tools}

A comprehensive treatment is essential for characterizing the heralded states properties. In order to find a desired parameter space for the state generation we use the fidelity, success probability, and $g^{(2)}_{\text{h}}$ as figures of merit for the heralded state. Further, we investigate the photon-number parity of the heralded states to be able to estimate the heralded state quality. 

\subsection{Fidelity}

The fidelity is the projection of the generated state onto the desired targeted photon-number state with $m$ photons, therefore, it represents the quality of the heralded state. This figure of merit is affected by the imperfect heralding and the losses in the signal detection. In the case of perfect overlap the fidelity equals to the unity. However, in the case of imperfect heralding one encounters $\mathcal{F}_{\text{h}}(\ket{m})< 1$, with $m=1,2,3$. Numerically, the maximum possible fidelity can be obtained from the loss-degraded photon-number distribution of the heralded state, equation~(\ref{eq:Lossy_statistics}), by extracting the corresponding photon-number contribution $\mathcal{F}_{\text{h}}(\ket{m}) = p^\text{s}_{m,L}\,$.

\subsection{Success probability}

Due to the probabilistic nature of the PDC process, there is finite probability that a certain target state can be created. The success of heralding a given state is determined by the probability that a specific measurement projection happens. The success probability is equivalent to the normalization factor in equation~(\ref{Eq:rho_signal}), that is, $ \mathcal{P}_{\text{h}} =	\Tr_{\text{s,i}}\left\{\hat{O}_k \hat{\rho}_{\text{s,i}}\right\}\,$.
Therefore, this probability also depends on all the relevant parameters of the POVM and the twin beams, but it is independent of the signal detection efficiency $\mu_{\text{s}}$.

\subsection{Normalized factorial moments of photon number}

The normalized m-th order factorial moment of photon number $g^{(m)}$ is a practical figure of merit for evaluating the photon-number content of faint light \cite{laiho2022measuring}. In comparison to the photon statistics these normalized moments are usually extracted in loss-independent manner. One can evaluate these moments for the heralded state in the signal beam path, denoted as $g^{(m)}_{\text{h}}$, while idler serves as a herald. In particular, we make use of the normalized second-order factorial moment of photon number for the heralded state, $g^{(2)}_{\text{h}}$, as a figure of merit. Numerically, the normalized factorial moments can be computed in terms of the photon-number distribution via
\begin{equation}
    g^{(m)}_{\text{h}} = \frac{\sum_{n}\frac{n!}{(n-m)!}p^\text{s}_n}{\left(\sum_{n}np^\text{s}_n\right)^m} = \frac{\sum_{n}\frac{n!}{(n-m)!}p^\text{s}_{n,L}}{\left(\sum_{n}np^\text{s}_{n,L}\right)^m}\, ,
\label{eq:high-orderCorrFunc}
\end{equation}
indicating that the value of $g^{(m)}_{\text{h}}$ remains the same for a single optical mode regardless whether the loss-degraded photon statistics from equation~(\ref{eq:Lossy_statistics}) or the lossless photon statistics in equation~(\ref{eq:heralded_statistics_2}) is employed \cite{barnett2002methods}. For more details see appendix~\ref{appx:normalized_high_order}.

\subsection{Photon-number parity of heralded states}

The expectation value for the photon-number parity can be directly evaluated from the photon statistics. Alternatively, if the mean-photon number of the heralded state is known as well, this figure of merit can be extracted via the normalized factorial moments of photon number that usually are measured in loss-independent fashion. In case the heralded state suffers from optical losses, the loss-corrected form of the heralded state mean photon number $\left<\hat{n}_{\text{s}}\right>$ is straightforward to determine \cite{laiho2022measuring}. This can be done by dividing the loss-degraded mean photon number of the heralded state $\left<\hat{n}_{\text{s},L}\right>$ with the detection efficiency in signal arm, that is  $\left<\hat{n}_{\text{s},L}\right>=\mu_{\text{s}} \left<\hat{n}_{\text{s}}\right>$ \cite{barnett2002methods}. 

Alternatively, also CAR offers an approximate approach for extracting the loss-corrected mean photon number of the heralded state. In section~\ref{sec:Results}, we estimate $\left<\hat{n}_{\text{s}}\right>$ in terms of the CAR value. Such finding also suggests the existence of a proper region in the CAR parameter to improve the generation of heralded number-photon states.

Numerically, the loss-degraded photon-number parity is gained by definition via \cite{barnett2002methods} 
\begin{equation}
    \left<\hat{\Pi}\right> = \sum_{n=0}^{\infty} (-1)^n p_{n,L}^\text{s} =  \sum_{m=0}^{\infty} \frac{g^{(m)}_{\text{h}}}{m!}\left(- 2 \left<\hat{n}_{\text{s},L}\right>\right)^m \, ,
\label{eq:parity}
\end{equation}
where $p_{n,L}^\text{s}$ is the loss-degraded photon-number distribution given in equation~(\ref{eq:Lossy_statistics}), $g^{(m)}_{\text{h}}$ is the normalized factorial moment described in equation~(\ref{eq:high-orderCorrFunc}), and $\left<\hat{n}_{\text{s},L}\right>$ is the loss-degraded mean-photon number of the heralded state in the signal beam. Therefore, losses in the signal arm can in principle be fixed by implementing the inverse of the loss-matrix for the evaluation of the loss-inverted photon statistics or using the loss-corrected mean-photon number of the heralded state $\left<\hat{n}_{\text{s}}\right>$. We note that care has to be taken when applying right hand side of equation~\ref{eq:parity} that the summation converges.


\section{Results}
\label{sec:Results}
\begin{figure}
    \centering
    \includegraphics[width=0.9\linewidth]{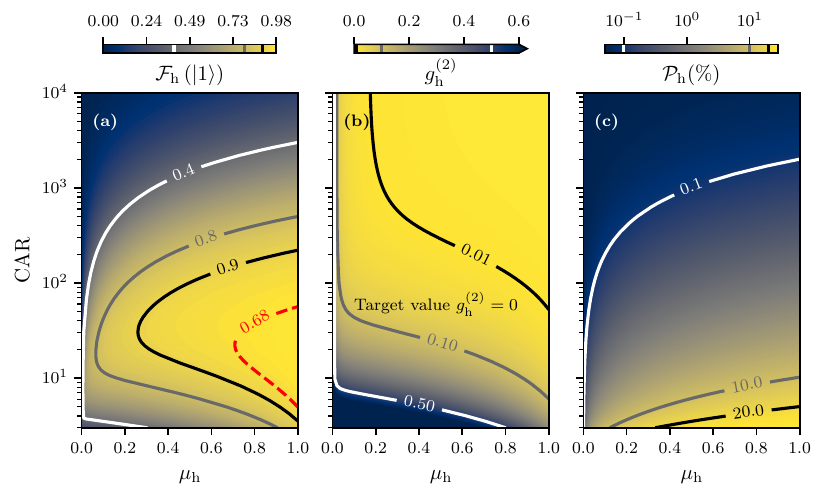}
    \caption{Properties of heralded single photons. We investigate (a) fidelity, (b) $g^{(2)}_{\text{h}}$ and (c) success probability in terms of the heralding efficiency $\mu_{\text{h}}$ and CAR value, which is inversely related to the twin beam mean photon number, with $\mu_{\text{s}}=1$ and other input parameters listed in table~\ref{Tab:parameters}. Additionally, the ticks for the contour values (white, gray, black) are marked in the colorbars. The red-dashed contour line corresponds to the fidelity with an efficiency in the signal arm of $\mu_{\text{s}}=0.7$.}
\label{fig:Fidelity_g2_prob_1}
\end{figure}

We simulate the characteristics of the heralded number states with the help of the analysis tools defined in section~(\ref{sec:Tools}). The states are generated in terms of the input parameters treated in section~(\ref{sec:Methods}). Apart from the fixed POVM parameters, these particular input parameters of the state preparation underpin to the CAR, and the heralding and detection efficiencies $\mu_{\text{h}}$ and $\mu_{\text{s}}$, respectively. In the following, we compute the fidelity, $g^{(2)}_{\text{h}}$, and success probability within a two-dimensional parameter space defined by the CAR and the heralding efficiency $\mu_{\text{h}}$. For extracting the fidelity we assume an ideal signal detection efficiency of $\mu_{\text{s}} = 1$, giving us the clearest picture of the achievable values in the sense of an upper limit. Besides, these numbers are not contaminated by multiphoton contributions, which may happen with other values of $\mu_{\text{s}}$. However, for the sake of clarity we include in the fidelity results a contour line for $\mu_{\text{s}}=0.7$, which represents an optimistic but more realistic value for experiments. We note that the other two used figures of merits are independent of it. Additionally, the photon-number parity is evaluated in a three-dimensional parameter space that also includes variations in the signal detection efficiency. In this way, we can identify the region for preparing the most desired state using readily accessible experimental parameters. 

We start by investigating the characteristics of the heralded single-photon state by setting $k=1$. In figure~\ref{fig:Fidelity_g2_prob_1} we present the results for its fidelity, $g^{(2)}_{\text{h}}$ and success probability. In the following, we depict for all investigated states the desired regions with bright yellow, while less favorable regions are shown in dark blue. A substantial region of the parameter space, bounded by the black curve in figure~\ref{fig:Fidelity_g2_prob_1}(a), exhibits a fidelity of $\mathcal{F}_{\text{h}}(\ket{1})\geq 0.90$ indicating the region of high overlap with the target state. The red-dashed contour line presents a fidelity of $\mathcal{F}_{\text{h}}(\ket{1})\geq 0.68$ with efficiency in the signal arm $\mu_{\text{s}}=0.7$. This depicts the degradation of the heralded state by such efficiency.

This high-fidelity region corresponds to heralding arm efficiencies $\mu_{\text{h}}$ ranging from around 0.3 to 1.0 and values of CAR approximately between 4 and 230. Even though we assume an ideal signal detection efficiency of $\mu_{\text{s}} = 1.0$, we can extract a minimum required heralding efficiency and upper and lower bounds for the value of CAR from our results. Outside this region, towards higher values of CAR the fidelity diminishes due to an increased contribution from the vacuum component, and towards lower values of CAR the fidelity decreases as multi-photon contributions become more prominent. Notably, higher heralding detection efficiencies consistently lead to improved fidelity values, as anticipated. The maximum fidelity of $\mathcal{F}_{\text{h}}(\ket{1}) \approx 0.98$ is achieved within a narrow CAR parameter range from around $15$ to $38$ under ideal conditions with no losses, specifically $\mu_{\text{h}} = 1$ and $\mu_{\text{s}} = 1$.

The $g^{(2)}_{\text{h}}$, ideally approaches zero for a heralded single-photon state. However, this is unattainable in practice due to dark counts and multi-photon contributions. As shown in figure~\ref{fig:Fidelity_g2_prob_1}(b), $g^{(2)}_{\text{h}}$ disappears at high values of CAR. Conversely, at low values of CAR, $g^{(2)}_{\text{h}}$ increases due to the simultaneous generation of multiple photon pairs. Typically, $g^{(2)}_{\text{h}}\leq 0.5$ is considered a boundary for discriminating single-photon sources \cite{grunwald2019effective}. However, at high values of CAR, the heralded state may have significant vacuum contributions, resulting in low-quality single-photon states, as evidenced by a reduction in fidelity. This highlights that a low $g^{(2)}_{\text{h}}$ value alone is insufficient to fully characterize the quality of a heralded single-photon state, especially at high values of CAR. 

Finally, we depict the success probability $\mathcal{P}_{\text{h}}$ in figure~\ref{fig:Fidelity_g2_prob_1}(c). This probability decreases at higher values of CAR due to a reduced generation rate of photon pairs. In the region of high heralding detection efficiencies the maximum success probability, $\mathcal{P}_{\text{h}}\approx29\%$, is achieved at low values of CAR. For ideal detection efficiencies, the success probability for states with fidelity above $0.90$ is approximately $26\%$. 

\begin{figure}
    \centering
    \includegraphics[]{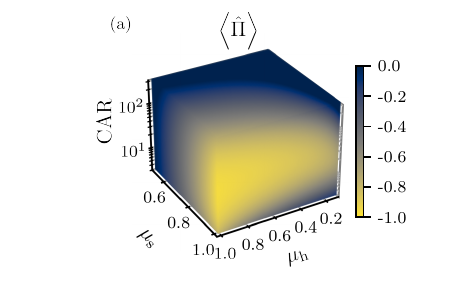}
    \includegraphics[]{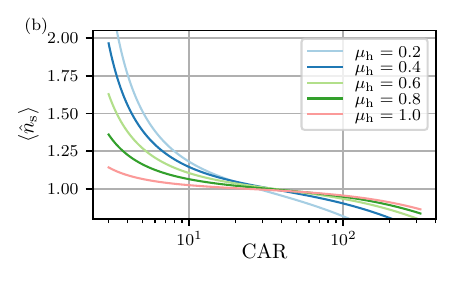}
    \caption{(a) Photon-number parity for a heralded single-photon state in terms of $\mu_{\text{h}}$, $\mu_{\text{s}}$ and CAR, with the input parameters listed in table~\ref{Tab:parameters}. (b) Loss-corrected mean-photon number of the heralded state as a function of CAR for different heralding detection efficiencies.}
\label{fig:PNP_1_3D}
\end{figure}

In figure~\ref{fig:PNP_1_3D}(a) we present the photon-number parity of the heralded single-photon state. For our target state this expectation takes the value $\left<\hat{\Pi}\right>=-1$. This region closely matches with the fidelity results in figure~\ref{fig:Fidelity_g2_prob_1}(a), underscoring the utility of the photon-number parity as a practical characterization tool. Unlike $g^{(2)}_{\text{h}}$, which provides only a limited insight into the state's quality, the photon-number parity gives an insightful characterization of the heralded state, offering a reliable alternative to the fidelity as a figure of merit. The effect of signal detection efficiency, $\mu_{\text{s}}$, is also evident in the results. A high detection efficiency of $\mu_{\text{s}}>80\%$ in the signal arm is required to effectively characterize the heralded state without loss inversion.

Indeed, the parity can be experimentally retrieved by using photon-correlation measurements, as depicted in equation~(\ref{eq:parity}). Additionally, instead of extracting the loss-degraded heralded state mean photon number $\left< \hat{n}_{\text{s},L}\right>$ and correcting it for losses, there is a connection between CAR and the loss-corrected mean-photon number of the heralded state $\left< \hat{n}_{\text{s}}\right>$, as shown in figure~\ref{fig:PNP_1_3D}(b). In the region of CAR around 10 and 100 one can reach $\left< \hat{n}_{\text{s}}\right>\approx 1$. This region also corresponds to the most favorable for the heralding of a single-photon state according to the results from parity, fidelity and $g^{(2)}_{\text{h}}$. 

Overall, the quality of the heralded single-photon state diminishes at high values of CAR due to the dark count probability and the increased vacuum contributions in the thermal distribution of the single-mode twin beams. Conversely, at low values of CAR, the quality is impaired by the growing multi-photon components in the heralded state. Altogether, the optimal heralded single-photon state is achieved at the value of CAR around $15$, with fidelity $\mathcal{F}_{\text{h}} (\ket{1})\approx0.98$, $g^{(2)}_{\text{h}}\approx0.04$, success probability $\mathcal{P}_{\text{h}}\approx6.8\%$ and photon-number parity $\left<\hat{\Pi}\right>\approx-0.95$.

\begin{figure}
    \centering
    \includegraphics[width=0.9\linewidth]{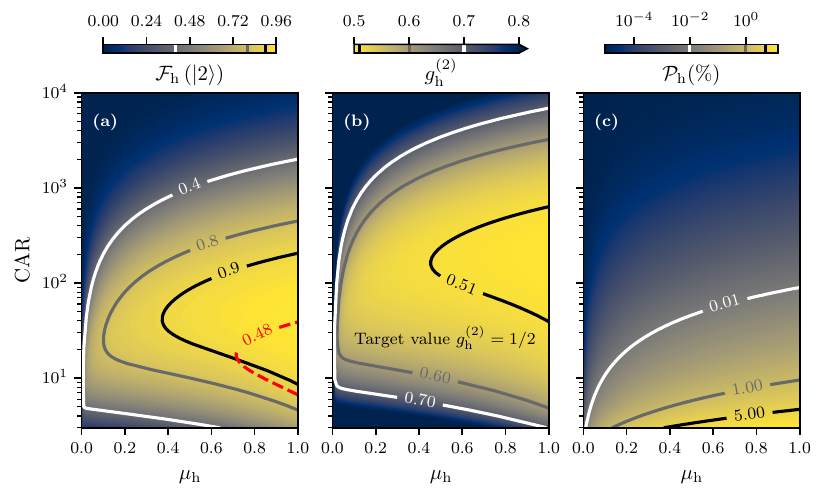}
    \caption{Properties of heralded two-photons. We investigate (a) fidelity, (b) $g^{(2)}_{\text{h}}$ and (c) success probability $\mathcal{P}_{\text{h}}$ in terms of the heralding efficiency $\mu_{\text{h}}$ and CAR value, with $\mu_{\text{s}}=1$ and other input parameters from table~\ref{Tab:parameters}. Additionally, the ticks for the contour values (white, gray, black) are marked in the colorbars. The red-dashed contour line corresponds to the fidelity with an efficiency in the signal arm of $\mu_{\text{s}}=0.7$.}
\label{fig:Fidelity_g2_prob_2}
\end{figure}

Second, we investigate the properties of the heralded two-photon state as a target state achieved in the simulation by setting $k = 2$, and present in figure~\ref{fig:Fidelity_g2_prob_2} its fidelity, $g^{(2)}_{\text{h}}$ and the success probability. Compared to the case in figure~\ref{fig:Fidelity_g2_prob_1}, the optimal region for generating a high-quality two-photon state is significantly smaller. Moreover, the success probability is notably lower with the highest success probability reaching approximately $14\%$ at low values of CAR. The fidelity for a detection with $\mu_{\text{s}}=0.7$ (red-dashed contour line) reduced the maximum fidelity to around $0.48$.

\begin{figure}[b]
    \centering
    \includegraphics[]{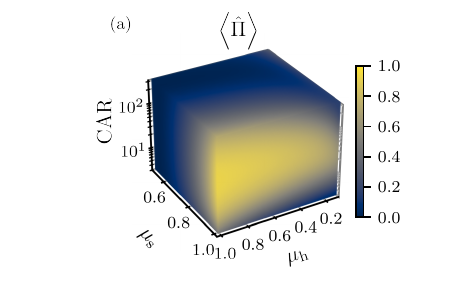}
    \includegraphics[]{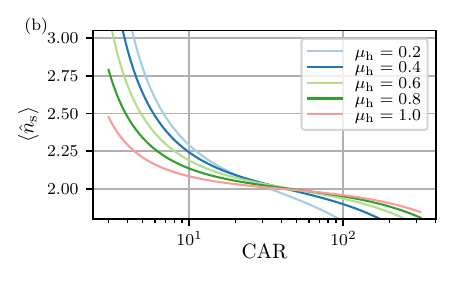}
    \caption{(a) Photon-number parity for a heralded two-photon state in terms of $\mu_{\text{h}}$, $\mu_{\text{s}}$ and CAR, with the input parameters listed in table~\ref{Tab:parameters}. (b)  Loss-corrected mean-photon number of the heralded state as a function of CAR for different heralding detection efficiencies.}
\label{fig:PNP_2_3D}
\end{figure}

Again, the photon-number parity presented in figure~\ref{fig:PNP_2_3D}(a) effectively encapsulates the information provided by both the fidelity and $g^{(2)}_{\text{h}}$, offering a comprehensive assessment of the heralded two-photon state’s quality. Ideally, the photon-number parity of the two-photon target state is $\left<\hat{\Pi}\right>=1$. Similar to the single-photon case, the vacuum contributions at high values of CAR reduce the state's purity, while multi-photon contributions dominate at low values of CAR, degrading the state’s quality. Once again, the heralding detection efficiency $\mu_{\text{h}}$ has less influence compared to the signal arm detection efficiency $\mu_{\text{s}}$, emphasizing the latter’s critical role. Finally, in figure~\ref{fig:PNP_2_3D}(b) a region of optimal heralded state mean photon number, $\left< \hat{n}_{\text{s}}\right>\approx 2$, is depicted. This region is found for an increased optimal CAR value compared to the case in figure~\ref{fig:PNP_1_3D}(b).

The photon-number parity reveals a relatively broad parameter space, where the heralded two-photon state maintains good quality. Notably, the purest generated state, with the highest success probability, is achieved at the value of CAR around $23$. This state exhibits a fidelity $\mathcal{F}_{\text{h}}(\ket{1})\approx 0.96$,  $g^{(2)}_{\text{h}}\approx0.52$, success probability $\mathcal{P}_{\text{h}}\approx0.15\%$ and photon-number parity value $\left<\hat{\Pi}\right>\approx 0.91$.

\begin{figure}[t]
    \centering
    \includegraphics[width=0.9\linewidth]{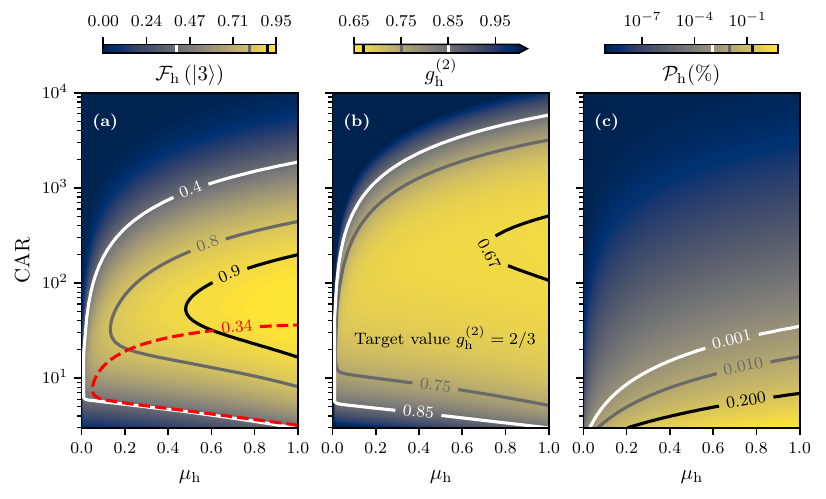}
    \caption{Properties of heralded three-photons. We investigate (a) fidelity, (b) $g^{(2)}_{\text{h}}$ and (c) success probability $\mathcal{P}_{\text{h}}$ in terms of the heralding efficiency $\mu_{\text{h}}$ and CAR value, with $\mu_{\text{s}}=1$ and other input parameters from table~\ref{Tab:parameters}. Additionally, the ticks for the contour values (white, gray, black) are marked in the colorbars. The red-dashed contour line corresponds to the fidelity with an efficiency in the signal arm of $\mu_{\text{s}}=0.7$.}
    \label{fig:Fidelity_g2_prob_3}
\end{figure}

Finally, we explore the properties of the heralded three-photon state as target state by setting $k=3$ and present in figure~\ref{fig:Fidelity_g2_prob_3} the results for the fidelity, $g^{(2)}_{\text{h}}$ and the success probability.
Compared to the cases of preparing single- and two-photon states, it becomes evident that the optimal parameter region for generating the desired state gets even smaller. For an efficiency of $\mu_{\text{s}}=0.7$, the state quality for heralded three-photon state possess a fidelity around $0.34$. Additionally, the success probability is further reduced, with the highest success probability reaching approximately $5.7\%$, for the low values of CAR.

\begin{figure}[t]
\centering
\includegraphics[]{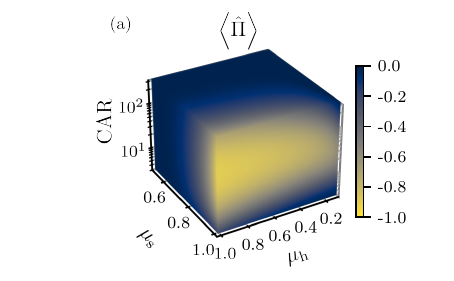}
\includegraphics[]{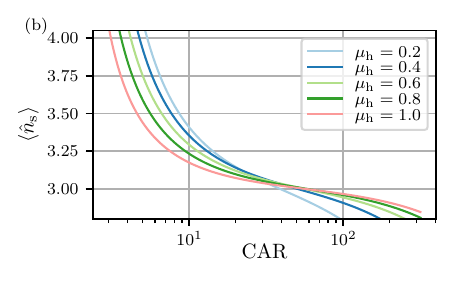}
\caption{(a) Photon-number parity for a heralded three-photon state in terms of $\mu_{\text{h}}$, $\mu_{\text{s}}$ and CAR, with the input parameters listed in table~\ref{Tab:parameters}. (b)  Loss-corrected mean-photon number of the heralded state as a function of CAR for different heralding detection efficiency.}
	\label{fig:PNP_3_3D}
\end{figure}

Similar to the cases of heralded single- and two-photon states, the effects of the vacuum and multi-photon contributions are observed also in the heralded three-photon state. For this purpose the value for the photon-number parity, which is presented in figure~\ref{fig:PNP_3_3D}(a), serves as a reliable indicator for identifying the region of high-quality heralded states. Ideally, the photon-number parity of the three-photon state takes the value $\left<\hat{\Pi}\right>=-1$. In figure~\ref{fig:PNP_3_3D}(b) we illustrate the mean photon number of the heralded state in terms of the value of CAR, indicating an optimal region for reaching the value $\left<\hat{n}_{\text{s}}\right>\approx 3$. A similar behavior of increased value of CAR is observed, as for the case in figure~\ref{fig:PNP_2_3D}(b). The purest generated state, with the highest success probability, occurs at the value of CAR around $43$, exhibiting a fidelity $\mathcal{F}_{\text{h}}(\ket{1})\approx0.95$, $g^{(2)}_{\text{h}}\approx0.675$, success probability $\mathcal{P}_{\text{h}}\approx5\times10^{-4}\%$, and a photon-number parity $\left<\hat{\Pi}\right>\approx-0.85$. Finally, we summarize in table~\ref{tab:High-Quality-State} the figures of merit for the heralded number states and best input parameter range in terms of CAR.

\begin{table}[b]
\centering
\begin{tabular}{cccc}
Figure of merit         & $\ket{1}$ & $\ket{2}$ & $\ket{3}$ \vspace{1mm} \\ 
\hline \vspace{-5mm}\\ \hline
CAR                     & $15$     & $23$      &  $43$  \\ 
$\mathcal{F}_{\text{h}}$         & $0.98$   & $0.96$    &  $0.95$   \\
$g^{(2)}_{\text{h}}$           & $0.04$   & $0.52$    &  $0.675$   \\
$\mathcal{P}_{\text{h}}~(\%)$  & $6.8$    & $0.15$    &  $5\times10^{-4}$   \\
$\left<\hat{\Pi}\right>$& $-0.95$  & $0.91$    & $-0.89$    \\
$\left< \hat{n}_{\text{s}}\right> (0.4\leq\mu_{\text{h}}\leq0.6)$    & $1.06-1.08$& $2.06-2.07$& $3.02-3.03$
\end{tabular}
\caption{Parameters for the state preparation together with those for describing the heralded state quality leading to characteristics closest to the target states.}
\label{tab:High-Quality-State}
\end{table}
\section{Conclusion}

The preparation of high quality heralded photon-number states plays a crucial role in quantum optics. The lack of true photon-number-resolving detection techniques at the few photon level restricts the quality of the state preparation and characterization. We studied the generation and characterization of heralded low photon-number states up to three photons prepared with heralding from single-mode twin beams generated via PDC. By analyzing the key experimental parameters such as the coincidences-to-accidentals ratio (CAR), heralding efficiency $(\mu_{\text{h}})$, and signal detection efficiency $(\mu_{\text{s}})$ we identified the trade-offs between success probability and state quality. Using figures of merits including the fidelity, normalized second-order factorial moment $\left(g^{(2)}_{\text{h}}\right)$, success probability and photon-number parity, we demonstrated how these state preparation parameters influence the quality and practical use of the heralded photon-number states.

Our results show that while fidelity and $g^{(2)}_{\text{h}}$ are widely used to characterize photon-number states, they have limitations. While the former needs complex loss-inversion techniques for photon statistics the latter can be used only for state classification. In contrast, the photon-number parity provides a comprehensive and experimentally accessible metric for assessing state quality, clearly showing the contamination from the undesired photon-number contributions.

Using a realizable heralding detector, we demonstrate that the optimal region for generating high-quality photon-number states depends on a delicate balance of experimental parameters. High values of CAR mitigate multi-photon contributions but increase the vacuum component and allow contamination through detector dark counts, while low values of CAR lead to excessive multi-photon contamination. Our results suggest that moderate levels of CAR are favorable, even for the heralding of a single-photon state. Additionally, detection efficiencies in both the heralding and signal arms significantly influence state quality. We identified regions in the input parameter space, where several analysis tools indicate high-quality state preparation, thus providing practical guidance for experimental implementations. 

In addition, our results point out that the desired value of CAR is increasing, when going towards a higher number state. This creates a challenge since at the same time the success probability is strongly reduced. Earlier studies have reported that there exist a trade-off between the success probability and state quality \cite{christ2012limits}. Our analysis can individually address the effect of the different experimental factors. Thus, we gain information of their interplay in a real experiment and can find the best range for each of these turning knobs. Including to the analysis of the fidelity and success probability also the analysis of $g^{(2)}_{\text{h}}$ and photon-number parity, we can compare the effect of the imperfections on each figure of-merit. Additionally, we provide means for the calibration of the heralded state mean photon number solely by the value of the CAR, which can be highly practical in an experiment.

Altogether, our study emphasizes the importance of the photon-number parity as a versatile and reliable tool for characterizing photon-number states, surpassing conventional metrics in practical utility and interpretive power. By identifying optimal regions in the experimental parameter space, this study contributes to the generation of high-quality heralded photon-number states, laying a foundation for their deployment in quantum optics tasks and contributing to the development of versatile photon-pair sources.


\section{Appendix}

\subsection{Derivation of the POVM presentation in the number state basis}
\label{appx:POVM}

The POVM for an array of $N$-detectors from which $k$ of them detect a photon is given by \cite{sperling2012true,sperling2014quantum}
\begin{equation}
    \hat{O}_k = :\binom{N}{k} \hat{\Theta}^{N-k} (\hat{\mathcal{I}}-\hat{\Theta})^k: \,, \hspace{1cm} \hat{\Theta}=\exp{-\frac{\mu_{\text{h}}\hat{n}+\nu}{N}} \, ,
\end{equation}
where $\hat{\mathcal{I}}$ is the identity operator, and $\hat{\Theta}$ takes into account the efficiency $\mu_{\text{h}}$ of the detectors and the dark count probability $\nu$. By expanding the binomial from the $k$-powered term,
\begin{equation}
    \hat{O}_k = :\binom{N}{k} \hat{\Theta}^{N-k} \sum_{m=0}^k \binom{k}{m} \hat{\mathcal{I}}^{k-m}(-\hat{\Theta})^m: \, .
\end{equation}

Then rearranging and replacing the $\hat{\Theta}$ operator gives, and noting that $\hat{\mathcal{I}}^j = \hat{\mathcal{I}}$,
\begin{equation}
    \hat{O}_k = \sum_{m=0}^k \binom{N}{k} \binom{k}{m}(-1)^m e^{-\frac{\nu}{N}(N+m-k)} :\exp{-\frac{\mu_{\text{h}} \hat{n}}{N}(N+m-k)}: \, .
\end{equation}
Now, by making used of the identity for ordered exponential operators \cite{barnett2002methods}
\begin{equation} 
 :\exp{\alpha \hat{n}} : = \sum_{n=0}^\infty (1+\alpha)^{n} \ket{n}\bra{n} \, ,
\end{equation}
and taking into account that the operator applies on the idler beam, it is possible to get the form presented in equation~(\ref{eq:O})
\begin{equation}
    \hat{O}_k = \sum_{m=0}^{k} \binom{N}{k}\binom{k}{m} (-1)^m e^{-\frac{\nu}{N}(N+m-k)}\sum^{\infty}_{n=0}\left(1-\frac{\mu_{\text{h}}}{N}(N+m-k)\right)^{n} \ket{n}_{\text{i~i}}\bra{n} \, .
\end{equation}

\subsection{Effect of the dark counts on the heralded single photon state}
\label{appx:darkcounts}
From equation~(\ref{eq:heralded_statistics_2}) one finds that for the heralded single-photon state,
\begin{equation}
    \frac{p_1^{\text{s}}}{p_0^{\text{s}}} = \frac{P_1(\bar{n})}{P_0(\bar{n})}\times \frac{\left( 1-\mu_{\text{h}}/N(N-1)\right)e^{\nu/N}-(1-\mu_{\text{h}})}{e^{\nu/N}-1} \, ,
\end{equation}
which corresponds to the ratio between single-photon and vacuum contribution in the heralded state photon statistics. In the limit of zero dark count probability, the ratio diverges to infinity, implying that the statistics is highly populated by the single-photon contribution as no vacuum contributions are presented. Similarly, as $\nu\rightarrow\infty$ the ratio goes to zero, meaning that the population in the vacuum is dominant. In the region of low dark counts, a first order expansion leads to
\begin{equation}
    \frac{p_1^{\text{s}}}{p_0^{\text{s}}} \propto \frac{P_1(\bar{n})}{P_0(\bar{n})}\times\frac{\mu_{\text{h}}}{\nu} \, ,
\end{equation}
in which the ratio of the heralding efficiency to the dark count probability governs the quality of the heralded single photon state by contributing the ratio between single-photon and vacuum contributions and being a dominant factor for reaching a high fidelity.

\subsection{Cross-correlation function}
\label{appx:Cross-correlation_function}
In a similar way to the normalized factorial moments, the high-order cross-correlation function can in general be computed as
\begin{equation}
    g^{(n,m)} = \frac{\Tr_{\text{s,i}}\left\{\hat{\rho}_{\text{s,i}}:(\hat{a}_{\text{i}}^\dagger \hat{a}_{\text{i}})^n: :(\hat{a}_{\text{s}}^\dagger \hat{a}_{\text{s}})^m:\right\}}{\left(\Tr_{\text{s,i}}\left\{\hat{\rho}_{\text{s,i}}\hat{a}_{\text{i}}^\dagger\hat{a}_{\text{i}}\right\}\right)^n \left(\Tr_{\text{s,i}}\left\{\hat{\rho}_{\text{s,i}}\hat{a}_{\text{s}}^\dagger\hat{a}_{\text{s}}\right\}\right)^m}\,,
\end{equation}
where $\hat{a}_{\text{s,i}}$ $(\hat{a}^\dagger_{\text{s,i}})$ is the generalized mode annihilation (creation) operator of the studied pulsed single-mode PDC process \cite{rohde2007spectral}. In terms of the joint photon-number statistics $P_{j,l}$,
\begin{equation}
    g^{(n,m)} =  \frac{\sum_{j,l}\frac{j!}{(j-n)!}\frac{l!}{(l-m)!}P_{j,l}}{\left(\sum_{j}jP_{j,l}\right)^n\left(\sum_{l}lP_{j,l}\right)^m} \, .
\label{eq:g_n_m_appx}
\end{equation}
Furthermore, as the single-mode twin beams are correlated in the photon-number, the joint photon-number statistics is diagonal as shown in equation~(\ref{eq:joint_statistics}), $P_{j,l}=P_j(\bar{n})$. Thus, equation~(\ref{eq:g_n_m_appx}) simplifies to
\begin{equation}
    g^{(n,m)} =  \frac{\sum_{j}\frac{j!}{(j-n)!}\frac{j!}{(j-m)!}P_{j}(\bar{n})}{\left(\sum_{j}jP_{j}(\bar{n})\right)^{n+m}} \, .
\label{eq:cross_correlated_SM_TWBS_appex}
\end{equation}
From this expression, the CAR value can be computed as in equation~(\ref{eq:CAR_smTWS}).

\subsection{Normalized high-order factorial moment}
\label{appx:normalized_high_order}
\begin{figure}
    \centering
    \includegraphics[width=0.3\linewidth]{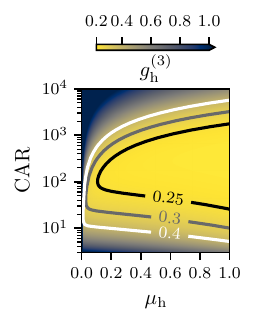}
    \caption{ $g^{(3)}_{\text{h}}$ value of heralded three-photons in terms of the heralding efficiency $\mu_{\text{h}}$ and CAR value, with $\mu_{\text{s}}=1$ and other input parameters from table~\ref{Tab:parameters}.}
\label{fig:Fidelity_g2_prob_3_g3}
\end{figure}

In general, the normalized high-order factorial moment for the heralded beam is retrieved from
\begin{equation}
    g^{(m)}_{\text{h}} = \frac{\Tr_{\text{s}}\left\{\hat{\rho}_{\text{s}}:(\hat{a}_{\text{s}}^\dagger \hat{a}_{\text{s}})^m:\right\}}{\left(\Tr_{\text{s}}\left \{\hat{\rho}_{\text{s}}\hat{a}_{\text{s}}^\dagger\hat{a}_{\text{s}}\right\}\right)^m}\, ,
\end{equation}
where $\hat{\rho}_{\text{s}}$ is the density matrix of the signal beam in equation~(\ref{Eq:rho_signal}). This form can be expressed in terms of the photon statistics of the signal beam, as shown in equation~(\ref{eq:high-orderCorrFunc}).

For completeness, we present in figure~\ref{fig:Fidelity_g2_prob_3_g3} the values of $g^{(3)}_{\text{h}}$ for the three-photon state. In this case the value for the target state is $g^{(3)}_{\text{h}}=2/9\approx0.22$. However, a more strict condition for the desired state preparation range was achieved via $g^{(2)}_{\text{h}}$, which is presented in figure~\ref{fig:Fidelity_g2_prob_3}.

\section*{Acknowledgments}
Special thanks to Thomas Dirmeier and Christoph Marquardt for helpful discussions about photon correlations.

\printbibliography
\end{document}